\documentclass[aps,prl,showpacs,twocolumn,superscriptaddress]{revtex4-1}
\usepackage{amssymb}
\usepackage{graphicx}
\usepackage{appendix}
\usepackage{dcolumn}
\usepackage{bm}
\usepackage{amsmath}
\usepackage{epstopdf}
\usepackage{xcolor}
\usepackage{float}
\usepackage[english]{babel}
\usepackage[colorlinks,linkcolor=magenta,citecolor=blue,urlcolor=blue]{hyperref}
\usepackage{subfigure}
\renewcommand{\Re}{\operatorname{Re}}
\begin{document}

\title{Eigenvalues restricted by Lyapunov exponent of eigenstates}
\author{Tong Liu}
%\thanks{t6tong@njupt.edu.cn}
\affiliation{Department of Applied Physics, School of Science, Nanjing University of Posts and Telecommunications, Nanjing 210003, China}
\author{Xu Xia}
\thanks{1069485295@qq.com}
\affiliation{Chern Institute of Mathematics and LPMC, Nankai University, Tianjin 300071, China}

\date{\today}

\begin{abstract}
 We point out that the Lyapunov exponent of the eigenstate places restrictions on the eigenvalue. Consequently, with regard to non-Hermitian systems, even without any symmetry, the non-conservative Hamiltonians can exhibit real spectra as long as Lyapunov exponents of eigenstates inhibit imaginary parts of eigenvalues. Our findings open up a new route to study non-Hermitian physics.
\end{abstract}

\pacs{71.23.An, 71.23.Ft, 05.70.Jk}
\maketitle

\section{Introduction.}
In physics, the observability of physical quantities demands that the measured eigenvalue must be a real number~\cite{Shankar}.
Therefore, quantum mechanics generally assumes that the observable quantity is the Hermitian operator.
However, with the breakthrough of quantum theory and the development of experimental technology~\cite{Lindblad,Photonic1,ultracold1,Bender1}, it has been proved that non-Hermitian systems can also host real eigenvalues and be observed experimentally~\cite{Xiao}. Currently, the study of real energy spectrum mainly focuses on two classes: (i) real energy spectrum of the symmetry protection, such as parity-time ($\mathcal{PT}$) symmetry~\cite{Bender2}; (ii) real energy spectrum induced by boundary conditions, such as open boundary condition in non-Hermitian skin effect~\cite{Yao1,Yao2}.

\begin{figure}
  \centering
  \includegraphics[width=0.5\textwidth]{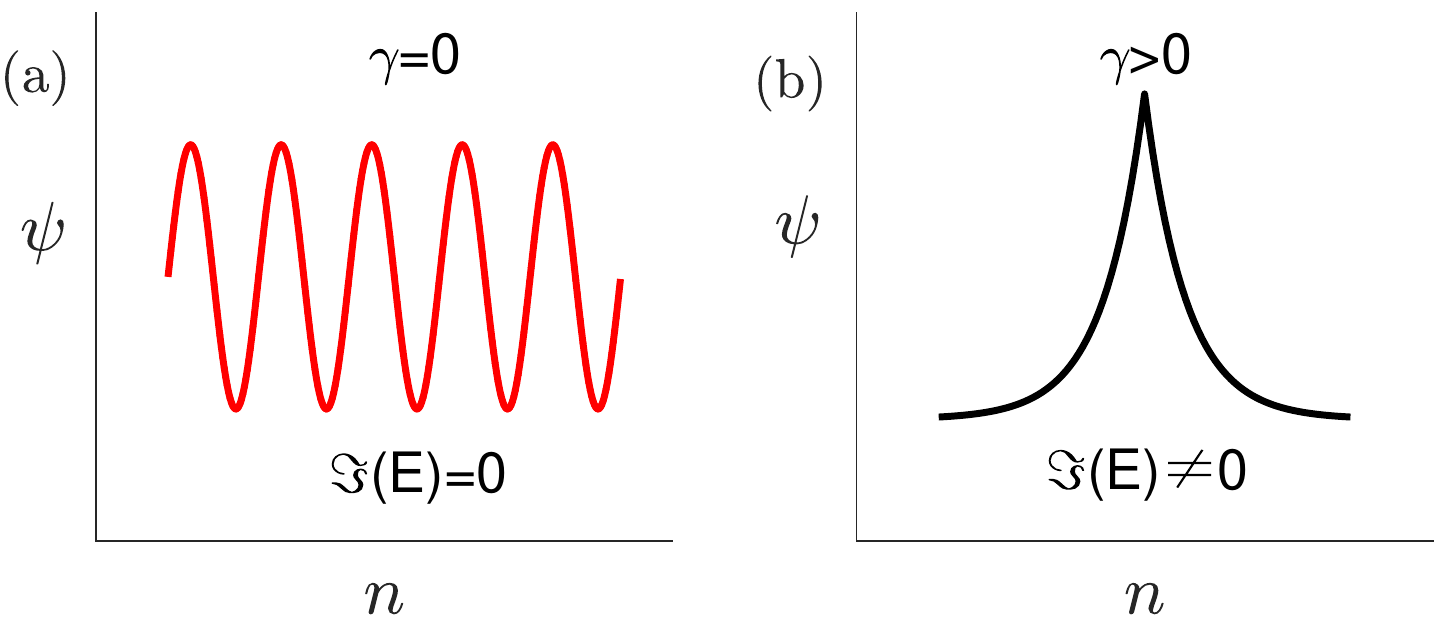}\\
  \caption{(Color online) (a)Cartoon diagram of an extended/critical eigenstate($\gamma=0$) versus $\Im(E)=0$. (b)Cartoon diagram of an localized eigenstate($\gamma>0$) versus $\Im(E)\neq0$.}
  \label{fig1}
\end{figure}

Thus, a question arises naturally: is there other physical mechanism for generating the real energy spectrum of non-Hermitian systems? In this work, we point out that the Lyapunov exponent (LE) of the eigenstate places restrictions on the eigenvalue, hence a non-Hermitian system can exhibit the real energy spectrum, independent of any symmetry and boundary conditions. Previous studies~\cite{Simon,Avila} have shown that the absolutely continuous spectrum of the Schr\"{o}dinger operators is completely determined by the LE, namely the absolutely continuous spectrum corresponds to the LE being zero. Thus, LE of eigenstates and eigenvalues establish a certain relationship. More explicitly, $\gamma$ of the eigenstate $\psi$ can always be expressed as a function of its eigenvalue $E$, namely
\begin{equation}\label{eq1}
\gamma\sim f(E),
\end{equation}
while the Lyapunov exponent $\gamma$ has only two possibilities, $\gamma=0$ for extended/critical eigenstates in Fig.~\ref{fig1}(a) and $\gamma>0$ for localized eigenstates in Fig.~\ref{fig1}(b). Hence $\gamma$ imposes restrictions on $E$, which may lead to the imaginary part $\Im(E)=0$, non-Hermitian systems can exhibit real energy spectrum.

To illustrate this point in detail, we analytically calculate a solvable model, namely the following non-Hermitian difference equation,
\begin{equation}
E \psi_n=\psi_{n+1}+\psi_{n-1}+V i \cot(\pi\alpha n) \psi_n,
\label{eq2}
\end{equation}
where $V$ is the complex potential strength, $E$ is the eigenvalue of systems, and $\psi_n$ is the amplitude of wave function at the $n$th lattice. We choose to unitize the nearest-neighbor hopping amplitude and a typical choice for irrational parameter $\alpha$ is $\alpha=(\sqrt{5}-1)/2$. Firstly, the complex potential $i \cot(\pi\alpha n)$ doesn't satisfy $V(n)=V^*(-n)$ in the discrete lattice, hence the model is independent of $\mathcal{PT}$ symmetry; secondly, the left and right hopping amplitudes are the same, hence the model is independent of non-Hermitian skin effect. Next we will calculate the explicit expressions of $\gamma$ and $E$, and give the condition that $E$ are real numbers.

\section{Lyapunov exponent and real energy.}
\begin{figure*}
    \begin{tabular}{cc}
    \subfigure[]{
    \begin{minipage}[t]{0.5\textwidth}
    \centering
    \includegraphics[width=1\textwidth]{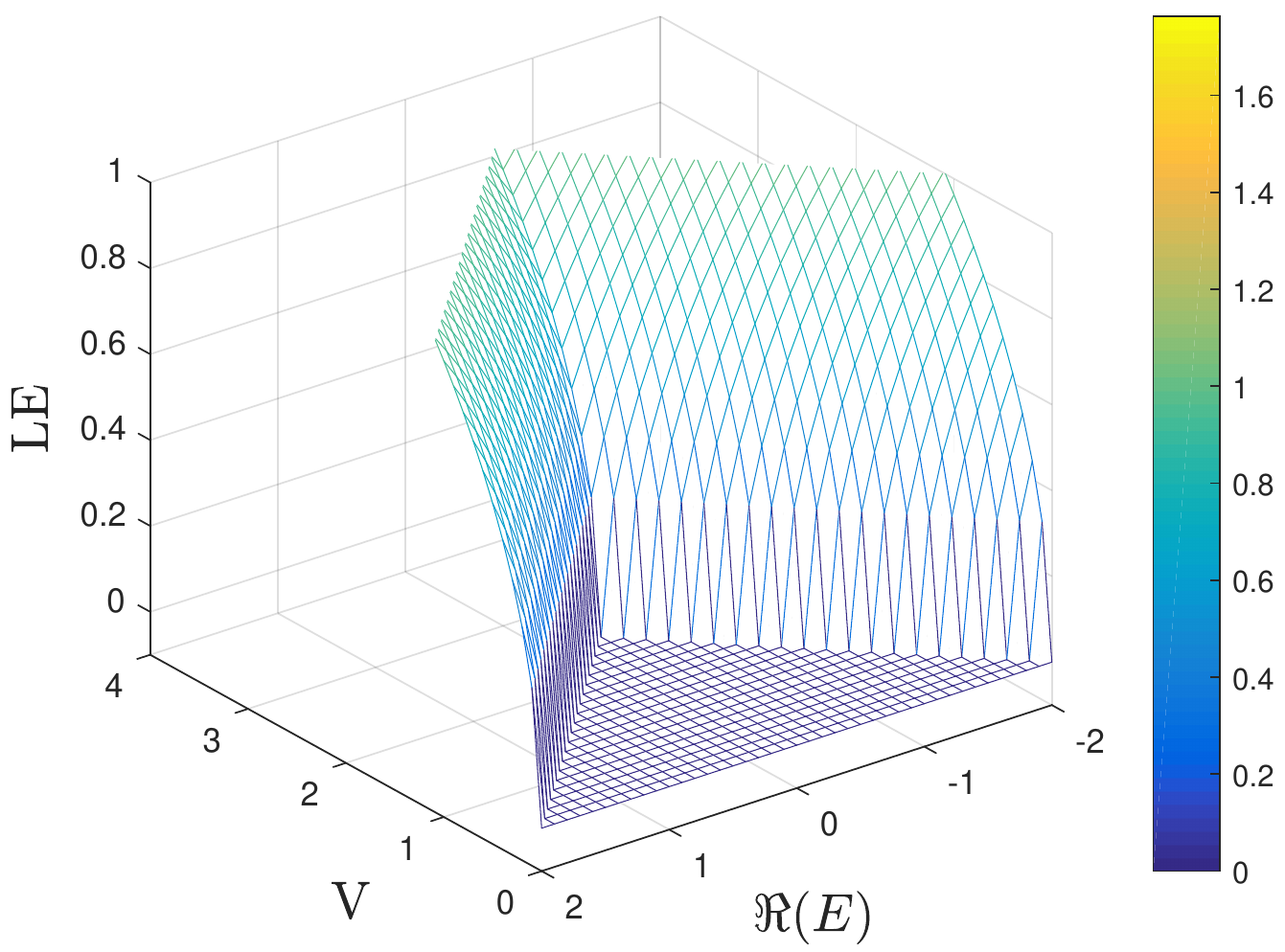}
    \end{minipage}}
   % \hspace{0.01\textwidth}
    \subfigure[]{
    \begin{minipage}[t]{0.5\textwidth}
    \centering
    \includegraphics[width=1\textwidth]{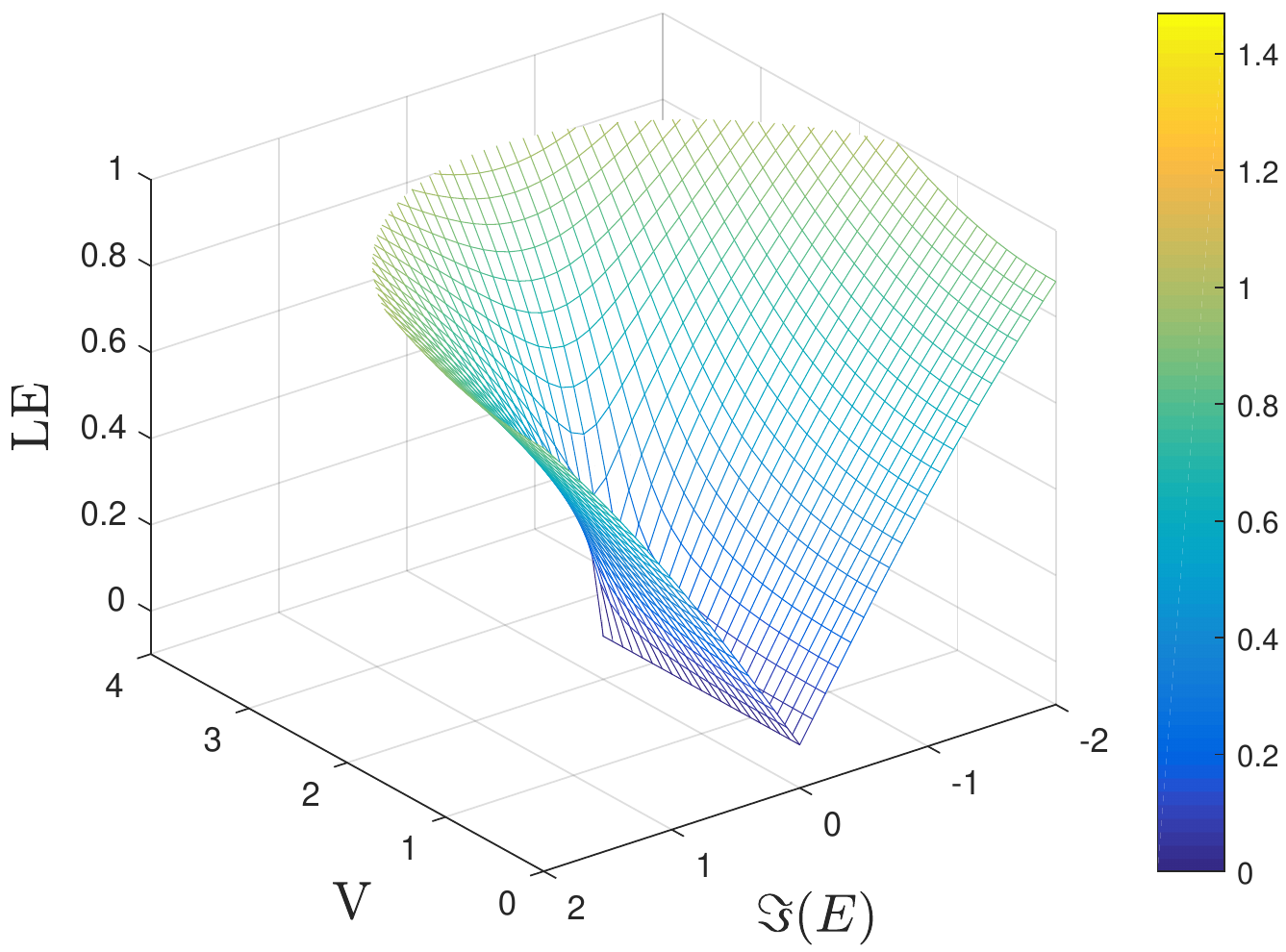}
    \end{minipage}}\\
    \end{tabular}
\caption{(Color online) (a) $\Re(E)$ versus LE as a function of the potential strength $V$. It clearly shows that when $\Re(E) \in [V-2, 2-V]$, $\gamma(E)=0$; otherwise $\gamma(E)>0$. (b) $\Im(E)$ versus LE as a function of the potential strength $V$. It clearly shows that when $\Im(E)=0$, $\gamma(E)=0$; otherwise $\gamma(E)>0$.}
\label{fig2}
\end{figure*}

The definition of LE has many different forms, the following form is adopted for a semi-infinite chain~\cite{Thouless},
\begin{equation}\label{eq3}
\gamma = - \lim_{n \rightarrow \infty} \frac{1}{n-n_0} \log \left| \frac{\psi_{n}}{\psi_{n_0}} \right|,
\end{equation}
where $\psi_{n_0}$ denotes the amplitude of wave function at the truncation of left end , $\psi_{n}$ denotes the amplitude of wave function at the infinity right end. Since Eq.~(\ref{eq2}) is a second-order difference equation, $\left| \frac{\psi_{n}}{\psi_{n_0}} \right|$ is a matrix form, which is very difficult to solve. Utilizing Avila's global theory~\cite{Avila}, we can get the explicit expression of LE,
\begin{equation}\label{eq4}
\begin{aligned}
\gamma(E) = \max\{&\operatorname{arcosh} \frac{\left| E+V+2\right|+\left| E+V-2\right|}{4},\\
&\operatorname{arcosh} \frac{\left|E-V+2\right|+\left|E-V-2\right|}{4}\} .
\end{aligned}
\end{equation}
Now, we get the explicit relation between $\gamma$ and $E$, a concrete example of Eq.~(\ref{eq1}). To visually illustrate the above result, a plot of $\gamma$ versus $E$ is shown in Fig.~\ref{fig2}.
Make $E=\Re(E)+i \Im(E)$, when $V\leq 2$, $\gamma(E)=0$ corresponds to the region $E \in [V-2, 2-V]$, while $\gamma(E)>0$ corresponds to the region $\Re(E)=0$ and $\Im(E)\in \mathbb{R}^{*}$. When $V> 2$, only $\gamma(E)>0$ exists, and $E$ belongs to the region $\Re(E)=0$ and $\Im(E)\in \mathbb{R}$. Here we need to emphasize that Eq.~(\ref{eq4}) only gives the value range of $E$ when $\gamma(E)=0$, and cannot prove that $E$ exactly corresponds to the eigenvalue of the system.

Hence we need prove the eigenvalues of Eq.~(\ref{eq2}) are exactly the value range of $E$ derived from $\gamma(E)$.
We fist introduce the Fourier transformation,
\begin{equation}\label{eq5}
f_\theta= \frac{1}{\sqrt{L}} \sum_{n=1}^L e^{i \theta n}\psi_n,
\end{equation}
thus the dual equation of Eq.~(\ref{eq2}) is given as,
\begin{equation}\label{eq6}
\begin{aligned}
&[\cos(\theta + 2\pi \alpha)+ V/2-E/2]f(\theta + 2\pi \alpha) \\
&= [-\cos(\theta - 2\pi \alpha)+ V/2+E/2]f(\theta - 2\pi \alpha).
\end{aligned}
\end{equation}
By Sarnark's method~\cite{Sarnak}, from Eq.~(\ref{eq6}), we can demonstrate that there are two kinds of eigenvalues of the system: (i) pure real numbers, namely $E \in [V-2, 2-V]$; (ii) pure imaginary numbers, the value range of imaginary part is the whole real axis. Consequently,
we demonstrates that when $\gamma(E)=0$, the non-Hermitian system hosts the real energy spectrum $E \in [V-2, 2-V]$, independent of any symmetry and boundary conditions, this is the central innovation of our work.

\section{Numerical verification}
\begin{figure}
  \centering
  \includegraphics[width=0.5\textwidth]{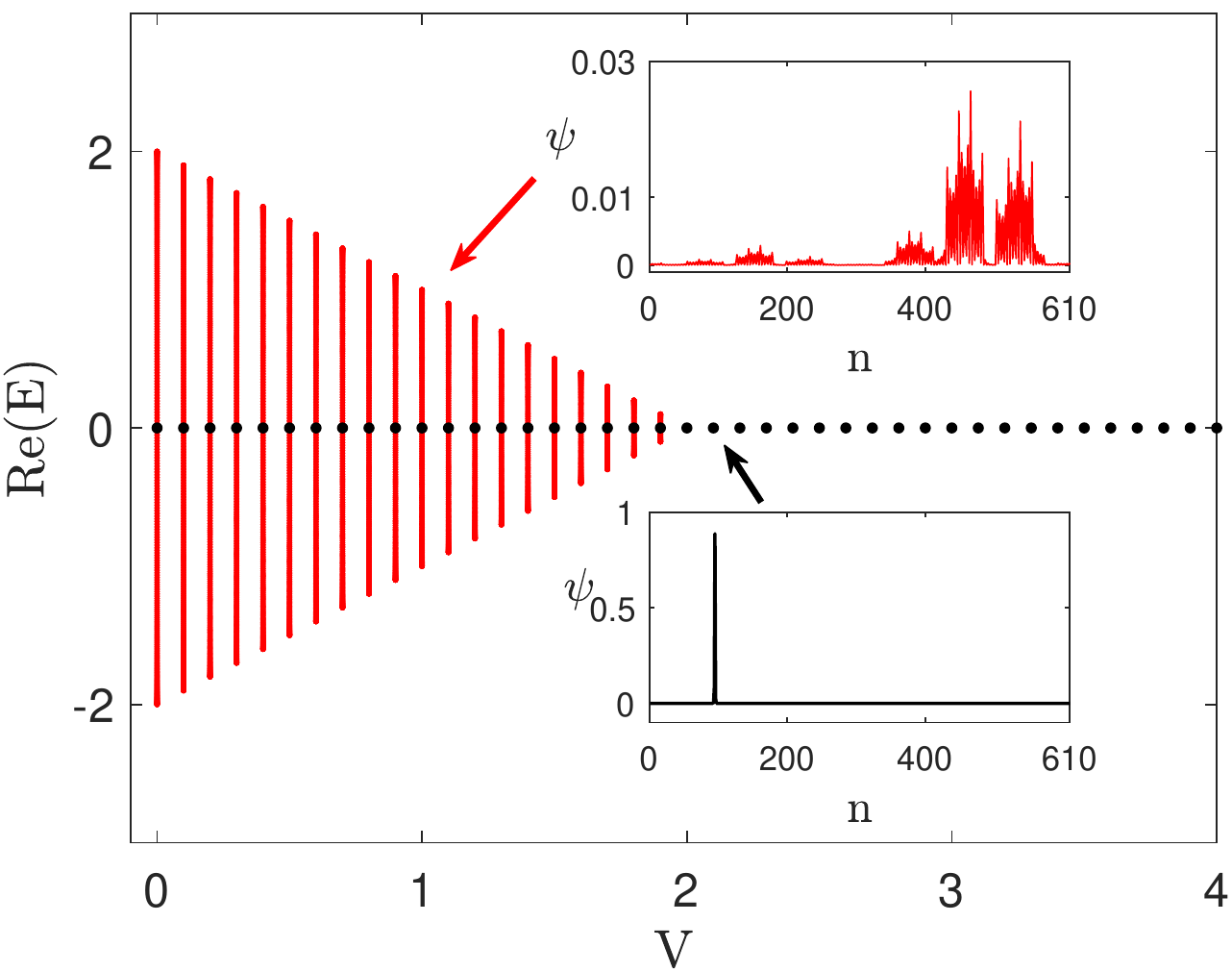}\\
  \caption{(Color online) The real part of eigenvalues $\Re(E)$ of Eq.~(\ref{eq2}) as a function of $V$. The total number of sites is set to be $L=610$. As shown in the figure, when $V\leq 2$, the system host the real energy spectrum $E \in [V-2, 2-V]$. In the inset, we plot the typical probability density, as expected, they are the critical state (red) corresponding to the pure real energy spectrum and the localized state (black) corresponding to the pure imaginary energy spectrum, respectively.}
  \label{fig3}
\end{figure}

To support the analytical result given above, we now present the numerical verification, namely directly diagonalize Eq.~(\ref{eq2}) to find the eigenvalues and eigenstates. With regard to the disordered system, the property of eigenstates not only can be characterized by LE, but also
can be measured by the inverse participation ratio (IPR) conveniently~\cite{Kohmoto}. For any given normalized eigenstate, the corresponding IPR is defined as $\text{IPR} =\sum_{n=1}^{L} \left|\psi_{n}\right|^{4}$, which measures the inverse of the number of sites being occupied by particles. The IPR of an extended state scales like $L^{-1}$ which becomes zero in the large $L$ limit, just as $\gamma=0$ in Fig.~\ref{fig1}(a). While for a localized state, the IPR is finite even in the large $L$ limit, just as $\gamma>0$ in Fig.~\ref{fig1}(b). In Fig.~\ref{fig3} we show the diagram of the eigenvalues $\Re(E)$ versus the complex potential strength $V$. The red eigenvalue curves denote the pure real energy spectrum hosting critical eigenstates, and the black circle dots denote the pure imaginary energy spectrum hosting localized eigenstates. These numerical results are completely consistent with the theoretical results derived by LE, which confirms the correctness of our theory.

\section{Conclusion.} In this work, we uncover a new class of physical systems with pure real energy spectrum, which is different from the known physical mechanism, namely independent of any symmetry and boundary conditions. Our research shows that this theory does not depend on any specific physical system, such as optics, cold atoms or classical circuits. As long as the LE of the eigenstate is determined, the eigenvalue of the system may have a pure real energy spectrum, which means that it has observable effects. $\gamma=0$ leads to the real eigenvalue in most cases, however, we need to emphasize that $\gamma=0$ is not a necessary condition for the eigenvalue to be a real number, and $\gamma>0$ may also produce a real eigenvalue, as long as $\Im(E)=0$. Therefore, there are still many academic gaps to be filled.
%%%%%%%%%%%%%%%%
\begin{acknowledgments}
This work was supported by the Natural Science Foundation of Jiangsu Province (Grant No. BK20200737), NUPTSF
(Grants No. NY220090 and No. NY220208), the National Nature Science Foundation of China (Grant No. 12074064), and
the Innovation Research Project of Jiangsu Province (Grant No. JSSCBS20210521). X.X. is supported by Nankai Zhide Foundation.
\end{acknowledgments}

%%%%%%%%%%%%%%%%%%%%%%%

\end{document}